\documentclass[preprint,12pt]{elsarticle}

\usepackage{amsmath}
\usepackage{amssymb}
\DeclareMathOperator\dif{d\!}

\usepackage{makecell}
\newcounter{bla}

\journal{Computer Physics Communications}

\begin{document}

\begin{frontmatter}



\title{mDCThermalC: A program for calculating thermal conductivity quickly and accurately}


\author{Tao Fan\corref{author}}
\author{Artem R. Oganov}

\cortext[author]{Corresponding author.\\\textit{E-mail address:} Tao.Fan@skoltech.ru}
\address{Skolkovo Institute of Science and Technology, 3 Nobel St., 121205 Moscow, Russia.}

\begin{abstract}
\textsf{mDCThermalC} is a program written in Python for computing lattice thermal conductivity of crystalline bulk materials using the modified %
Debye-Callaway model. Building upon the traditional Debye-Callaway theory, the modified model obtains the lattice thermal conductivity by averaging the contributions from acoustic and optical branches based on their specific heat. The only inputs of this program are the phonon spectrum, phonon velocity and Grüneisen parameter, all of which can be calculated using third-party \textit{ab initio} packages, making the method fully parameter-free. This leads to a fast and accurate evaluation and enables high-throughput calculations of lattice thermal conductivity even in large and complex systems. In addition, this program calculates the specific heat and phonon relaxation times for different scattering processes, which will be beneficial for understanding the phonon transfer behavior. 

\end{abstract}

\begin{keyword}
lattice thermal conductivity \sep Debye-Callaway model \sep phonon relaxation time

\end{keyword}

\end{frontmatter}



{\bf PROGRAM SUMMARY}

\begin{small}
\noindent
{\em Program Title: \upshape{mDCThermalC}}                                          \\
{\em Licensing provisions: \upshape{GPLv3}}                                   \\
{\em Programming language: \upshape{Python}}                                   \\
{\em External routines/libraries: \upshape{Numpy, Scipy, spglib, pymatgen}}                                   \\
{\em Nature of problem: }\\
The calculation of thermal conductivity from first principles method with an anharmonic approximation requires a large number of calculations to construct the third-order force constants matrix, which could be prohibitively long time. \\
{\em Solution method: }\\
modified Debye-Callaway model, only phonon spectrum, phonon velocity and Grüneisen parameter are needed. The acoustic branch and optic branch are both considered to obtain the final lattice thermal conductivity. 
%
\end{small}

\section{Introduction}
\label{Intro}
Studies of thermal conductivity and its behavior under extreme conditions play an important role in various technological and scientific applications, including thermoelectricity, heat management, and investigations of the deep Earth (mantle and core of the earth). Calculations of thermal conductivity that use as input only the basic information like crystal structure and do not require any other parameters obtained from experiments, while maintaining a sufficient accuracy, could be especially helpful in such research. Until recently, there already been several approaches which could fulfill this objective such as relaxation time approximation (RTA), iterative solution of the Boltzmann transport equation (BTE) and  \textit{ab initio} molecule dynamics. These methods are used in such software tools as  Phono3py\cite{togo2015}, ShengBTE\cite{li2014shengbte}, LAMMPS. However, all these approaches need large amounts of computing resources. Addressing these issues, here we present a small but robust computer program that calculates the lattice thermal conductivity, which is the phonon potion of the thermal conductivity of bulk crystalline materials. The program works fast and shows good accuracy.

Lattice thermal conductivity can be calculated directly using the temperature gradient from the nonequilibrium molecular dynamics (MD) simulation at a given heat current\cite{schelling2002, mcgaughey2006} or from the equilibrium MD simulations using the Green-Kubo method\cite{mcquarrie2000}. However, the MD simulations need a large unit cell to take into account the finite size effect and a long simulation time to converge the autocorrelation function. The second method to calculate lattice thermal conductivity is solving the phonon Boltzmann transport equation (PBTE), for which the most conventional way is to use the relaxation time approximation (RTA) along with the Debye approximation. Based on the RTA, a full iterative solution to the PBTE was developed\cite{broido2007}. However, these calculations need the second- and third-order interatomic force constants. Although density functional theory (DFT) is a convenient tool for accurately calculating the interatomic interactions in many cases, obtaining the third-order force constants used in the description of anharmonicity in phonon-phonon processes is still time-consuming. Another way to calculate lattice thermal conductivity $\kappa_L$ from first principles is the Debye-Callaway model.

In 1959, Callaway proposed a solution for the PBTE based on three assumptions\cite{callaway1959}: first, only four scattering mechanisms are considered, including point impurities (isotopes disorder), normal three-phonon processes, Umklapp processes and boundary scattering; second, all phonon scattering processes can be represented by frequency-dependent relaxation times; third, the crystal vibration spectrum is isotropic and dispersion-free. Based on this model, $\kappa_L$ of germanium was calculated in the temperature range of 2 K to 100 K. The results showed reasonable agreement with the experiments for both normal and single-isotope material. Asen-Palmer\cite{asen1997} modified the Debye-Callaway model by accounting for the contributions of longitudinal and transverse acoustic branches differently. In addition, this approach uses six freely adjustable parameters for longitudinal and transverse modes in order to include the anharmonic effect and contributions from the boundary and isotope scattering. Unfortunately, these models lack the predictive power since they incorporate parameters that are either fitted to experimental data or freely adjustable. Morelli \textit{et al.}\cite{morelli2002} modeled lattice thermal conductivity and isotope effect in Ge, Si, and diamond using an approach similar to that of Asen-Palmer. However, they used the known phonon dispersion relaitons of these crystals to derive all the necessary parameters except the Grüneisen parameter. Recently, Zhang\cite{zhang2016} developed a first-principles Debye-Callaway approach, where all the parameters (i.e., the Debye temperature $\Theta$, phonon velocity $\nu$ and Grüneisen parameter $\gamma$) can be directly calculated from the vibrational properties of compounds within the quasi-harmonic approximation.

In this paper we present a software package, \emph{mDCThermalC}, for calculating lattice thermal conductivity from the phonon dispersion curves obtained \textit{ab initio}. However, unlike the traditional Debye-Callaway model which only considers acoustic branches, our modified method also accounts for the contribution from optical branches, which could be as large as 20\% to 50\% at high temperatures, according to Slack\cite{slack1979}. We obtain the lattice thermal conductivity by averaging the contributions from the acoustic branches and optical branches based on their specific heat $c_v$. In addition, our software package can calculate the phonon-phonon relaxation times for different scattering mechanisms (including \textit{N} process, \textit{U} process and isotope scattering), which is important for the deep analysis and understanding of the heat transfer behavior.

The paper is structured as follows. In Section 2 we introduce the mathematical formalism and methodological choices behind mDCThermalC. Section3 includes testing examples for three different systems. We review our main conclusions and discuss some future directions for development in Section 4.

\section{Methodology}

\subsection{mDCThermalC workflow}

The software presented in this paper, enables approximate and highly efficient lattice thermal conductivity calculation from first principles. The main improvement of our method compared with Callaway model before is that the contribution from the optical branches is included and all the necessary physical quantities including Debye temperature, phonon velocity and Grüneisen parameter are obtained by first-principles calculations. The only inputs needed by mDCThermalC are phonon dispersion curves, phonon velocity and mode Grüneisen parameters. Our program relies completely on external tools (for now, it is Phonopy\cite{togo2008}) to generate them.

\begin{figure}[h]
\centering
\includegraphics[width=0.6\textwidth]{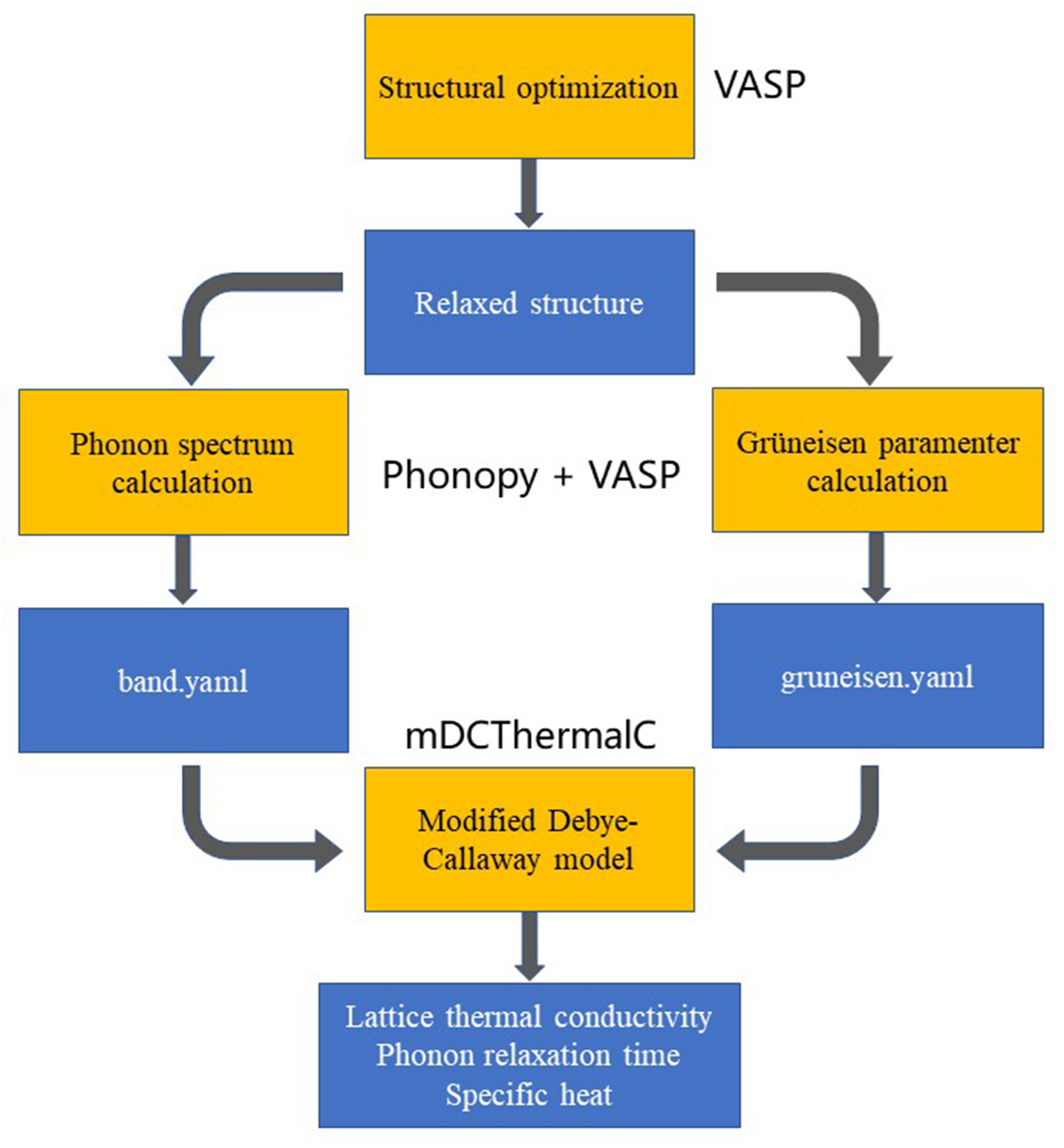} %

\caption{The workflow of lattice thermal conductivity calculation using mDCThermalC. Gold boxes represent steps of the calculation, blue boxes for the results of these steps, and computer programs are denoted as black text outside of the box.}\label{fig:1}
\end{figure}

We recommend to use VASP\cite{kresse1996software} combined with Phonopy to do the DFT calculations. A summary of our workflow, starting with structural optimization of each candidate compound and ending with the calculation of $\kappa_L$, is shown in Fig.\ref{fig:1}. 

\subsection{Modified Debye-Callaway model}

Following the approach used by Morelli\cite{morelli2002}, we update the formalism used to calculate lattice thermal conductivity. The total lattice thermal conductivity $\kappa$ is the weighted average of the acoustic branches (one longitudinal $\kappa_{LA}$, two transverse $\kappa_{TA}$ and $\kappa_{TA'}$ ) and one pseudo-optic branch ($\kappa_O$):
\begin{equation}\label{eq1}
\kappa = \frac{c^{aco}_V}{c^{aco}_V + c^{opt}_V} \times \frac{\kappa_{LA} + \kappa_{TA} + \kappa_{TA'}}{3}
              + \frac{c^{aco}_V}{c^{aco}_V + c^{opt}_V} \times \kappa_{O}
\end{equation}
where $\kappa_i = \kappa_{i1} + \kappa_{i2}$, with $i$ denoting $LA$, $TA$, $TA’$ and optic modes $O$. $c^{aco}_V$ and $c^{opt}_V$ are specific heat of acoustic and optical branches, which will be explained later. The partial conductivity $\kappa_{i1}$ and $\kappa_{i2}$ are the common Debye-Callaway terms:
\begin{subequations}
\begin{align}
\kappa_{i1} &= \frac 13 C_i T^3 \int_0^{\theta_i/T} \frac {\tau_C^i(x)x^4\mathrm{e}^x}{(\mathrm{e}^x - 1)^2}\dif x    \\
\kappa_{i2} &= \frac 13 C_i T^3 \frac {\int_0^{\theta_i/T} \frac {\tau_C^i(x)x^4\mathrm{e}^x}{(\mathrm{e}^x - 1)^2}\dif x} 
                         {\int_0^{\theta_i/T} \frac {\tau_C^i(x)x^4\mathrm{e}^x}{\tau_N^i(x)\tau_R^i(x)(\mathrm{e}^x - 1)^2}\dif x}
\end{align}
\end{subequations}
In these expressions, $\theta_i$ is the Debye temperature for each phonon branch, $C_i = k_B^4/(2\mathrm{\pi}^2\hbar^3\nu_i$) and $x = \hbar\omega/k_BT$, where $\hbar$ is the Planck constant, $k_B$ is the Boltzmann constant, $\omega$ is the phonon frequency, and $\nu_i$ is the phonon velocity for each branch; $(\tau_N^i)^{-1}$ is the scattering rate of the normal phonon process,  $(\tau_R^i)^{-1}$ is the total scattering rate of all the resistive scattering processes, and $(\tau_C^i)^{-1} = (\tau_N^i)^{-1} + (\tau_R^i)^{-1}$. According to Callaway, $(\tau_R^i)^{-1}$ should equal the sum of the scattering rates of the phonon-phonon Umklapp scattering, isotope point defect scattering, and scattering from the crystal boundary. In our model, only the Umklapp scattering and isotope scattering are considered, so that $(\tau_R^i)^{-1} = (\tau_U^i)^{-1} + (\tau_I^i)^{-1}$. For most practical applications like thermoelectricity, where the temperature is usually above 300 K, it is reasonable to omit the boundary scattering because it becomes significant only at very low temperatures, usually dozens of Kelvins. 

\subsubsection{Phonon-phonon normal scattering}

Although the normal phonon scattering is not a resistive process, it can redistribute the momentum and energy among phonons and influence other resistive scattering processes (such as the Umklapp scattering). Following the approach of Asen-Palmer\cite{asen1997}, the appropriate forms for longitudinal and transverse acoustic phonons are
\begin{subequations}
\begin{align}
[\tau_N^L(x)]^{-1} &= B_N^L(\frac{k_B}{\hbar})^2x^2T^5    
\intertext{and}
[\tau_N^T(x)]^{-1} &= B_N^T(\frac{k_B}{\hbar})^2xT^5
\end{align}
\end{subequations}
with the magnitudes $B_N$ depending on the phonon velocity $\nu$ and Grüneisen parameter $\gamma$,
\begin{subequations}
\begin{align}
B_N^L &= \frac{k_B^3\gamma_L^2V}{M\hbar^2\nu_L^5}    
\intertext{and}
B_N^T &= \frac{k_B^4\gamma_T^2V}{M\hbar^3\nu_T^5}
\end{align}
\end{subequations}
where $M$ is the average mass per atom in the crystal and $V$ is the volume per atom. A more general case and further discussion are included in the Appendix of Ref.\cite{morelli2002}. For the optical branch, we assume the same formula as for the longitudinal acoustic branch.

\subsubsection{Phonon-phonon Umklapp scattering}

The phonon-phonon Umklapp processes dominate at high temperatures, following an exponential behavior. According to Morelli\cite{morelli2002}, the Umklapp scattering rate for longitudinal and transverse acoustic phonons is:
\begin{subequations}
\begin{align}
[\tau_U^i(x)]^{-1} &= B_U^i(\frac{k_B}{\hbar})^2x^2T^3\mathrm{e}^{-\theta_i/3T}
\intertext{where}
B_U^i &= \frac{\hbar\gamma_i^2}{M\nu_i^2\theta_i}
\end{align}
\end{subequations}
The Umklapp scattering rate thus depends on Debye temperature, phonon velocity, and Grüneisen parameter of each branch. Again, we assume the optical branch to be described by the same formula as the longitudinal acoustic branch.

\subsubsection{Phonon-isotope scattering}

According to Klemens\cite{klemens1955}, the scattering rate of mass fluctuation due to the presence of isotopes should take the form
\begin{equation}
[\tau_I^i(x)]^{-1} = \frac{Vk_B^4\Gamma}{4\mathrm{\pi}\hbar^4\nu_i^3}x^4T^4
\end{equation}
Therefore, the isotope scattering rate also depends on the phonon velocity. The mass fluctuation phonon scattering parameter $\Gamma$ for a single element composed of several naturally occurring isotopes is 
\begin{subequations}
\begin{align}
\Gamma &= \sum_i c_i [\frac{m_i - \bar{m}}{ \bar{m}}]^2
\intertext{where}
 \bar{m} &= \sum_i c_i m_i
\end{align}
\end{subequations}
$m_i$ is the atomic mass of the \textit{i}th isotope and $c_i$ is the fractional atomic natural abundance. For a compound including $N$ different elements,
\begin{equation}
\Gamma(AB\dots) = N\left[ \left( \frac{M_A}{M_A+M_B+\dots} \right)^2\Gamma(A) + \left( \frac{M_B}{M_A+M_B+\dots} \right)^2\Gamma(B) + \dots \right] 
\end{equation}
where $M_i (i = A, B, \dots)$ denotes the average atomic mass of element $i$.

\subsubsection{Specific heat}

The specific heat is usually calculated using Debye model, which is only suitable for the acoustic branches. For structures whose primitive cell contains more than one atom $(p > 1)$, a more accurate method would be using the Debye model for the acoustic branches, while approximating the optical branches by the Einstein model. Then, the specific heat is 
\begin{subequations}
\begin{align}
c_V^{aco} &= 3\frac {N}{V} k_B f_D\left(\frac {\Theta_D}{T} \right) \\
c_V^{opt} &= (3p-3)\frac {N}{V} k_B f_E\left(\frac {\Theta_E}{T} \right) \\
f_D(x) &= \frac {3}{x^3} \int_0^x \frac{y^4\mathrm{e}^y\dif{y}}{(\mathrm{e}-1)^2}
\intertext{and}
f_E(x) &= x^2 \frac {\mathrm{e}^x}{(\mathrm{e}-1)^2}
\end{align}
\end{subequations}
where $\Theta_D$ is the Debye temperature, $\Theta_E$ is the Einstein temperature, $N$ is the number of primitive cells, $f_D$ and $f_E$ are the Debye function and Einstein function respectively.

\subsubsection{Debye temperature, Grüneisen parameter, and phonon velocity}

The lattice thermal conductivity is a function of the Debye temperature $\theta_i$, Grüneisen parameter $\gamma_i$ and phonon velocity $\nu_i$ of each phonon branch (Eqs. (1) - (9)), the parameters that can be readily obtained from an external software such as Phonopy.

For Debye temperature, we select the highest frequency of each branch to calculate $\theta_i$:
\begin{equation}
\theta_i = \frac {\hbar\omega_i^{max}}{k_B}
\end{equation}
The Debye temperature $\Theta_D$ in Eq.(9a) could, in principle, calculated from the specific heat at low temperatures. However, this would be an average of the whole spectrum, including both acoustic and optical branches. Here we need the contribution from the acoustic branches only. Thus, we select the maximum $\theta_i$ among the three acoustic branches as this $\Theta_D$.

The phonon velocity and Grüneisen parameter of each branch are calculated using a two-step averaging. The results of the Phonopy calculation of the phonon velocity and Grüneisen parameter are a function of band index $i$ and wavevector $\mathbf{q}$, namely $\nu(i,\mathbf{q})$ and $\gamma(i,\mathbf{q})$. The first average is taken within each high symmetry path of the same branch:
\begin{subequations}
\begin{align}
\nu(i,j) &= \overline{\nu(i,\mathbf{q})} \\
\gamma^2(i,j) &=  \overline{[\gamma(i,\mathbf{q})]^2}
\end{align}
\end{subequations}
where $j$ denotes a different high symmetry path in each branch. The second average is taken of these high symmetry paths:
\begin{subequations}
\begin{align}
\nu_i &= \frac {\sum_j m_j \nu(i,j)}{\sum_j m_j} \\
\gamma_i &= \sqrt{\frac{\sum_j m_j \gamma^2 (i,j)}{\sum_j m_j}}
\end{align}
\end{subequations}
where $m_j$ is the multiplicity of each high symmetry path, a value related with the symmetry of a structure.

We did not use all the optical branches in the thermal conductivity calculations. Instead, to correspond with the Einstein model and treat the optical branches' contribution as a correction to the original Callaway model, we use a “pseudo-optic” branch, which is an average of the optic branches. $\Theta_E$ in Eq. (9b) is the characteristic Einstein temperatrue of this "pseudo-optical" branch. 

\subsection{First-principles calculation setting}

All the first-principles calculations were performed using the VASP code with the projected augmented wave (PAW) potential\cite{kresse1996efficient, kresse1996efficiency}. The exchange-correlation energy was approximated by the PBE-GGA functional\cite{perdew1996}. For structural relaxation, the kinetic energy cutoff was set to 600 eV and the Brillouin zone was sampled using $\Gamma$-centered Monkhorst-Pack (MP) meshes\cite{monkhorst1976} with reciprocal space resolution of $2\mathrm{\pi}\times0.03\mathrm{\AA}^{-1}$. All structures were fully optimized until the total energy difference between the consecutive cycles was less than $10^{-7}$ eV and the maximum Hellmann-Feynman force was less than $10^{-3}$ eV/\AA. For the phonon calculation, a $2\times2\times2$ supercell was built and a $\Gamma$-centered mesh with the resolution of $2\mathrm{\pi}\times0.03\mathrm{\AA}^{-1}$ was used. The DFPT\cite{gonze1997} method was applied to obtain the second-order interatomic force constants (IFCs). To get Grüneisen parameters, three phonon calculations have to be run: one  at the equilibrium volume, the other two at slightly smaller (-0.4\%) and larger volume (+0.4\%) volumes. Especially, for SnSe, the van der Waals correction (IVDW=11) was included because the substance has a layered structure.

\section{Examples}

We illustrate the capabilities of mDCThermalC by performing calculations for three prototypical systems: (1) carbon as a diamond, an insulator with very high thermal conductivity; (2) silicon, the most studied semiconductor with normal thermal conductivity value; (3) tin selenide (SnSe), a famous thermoelectric material with the highest figure of merit achieved in its single crystal, mainly because of its very low thermal conductivity.

\subsection{Carbon (diamond)}

The phonon spectrum of diamond is shown in Fig. 2a. The lattice thermal conductivity of diamond calculated in this work is shown in Fig. 2b in comparison with the measurement results obtained in two studies\cite{onn1992,wei1993} and the calculation using the full \textit{ab initio} method\cite{ward2009}. The calculated Debye temperature, phonon velocity and Grüneisen parameter of diamond are listed in Table \ref{table:1}. We compared these parameters with previous work\cite{morelli2002} to make sure these input parameters are correct.

High lattice thermal conductivity of diamond is mainly caused by its very high phonon velocity. In general, our method tends to overestimate the value of the lattice thermal conductivity at low temperatures. For example, $\kappa_L$ obtained in our calculations at 300 K is 3180 $\mathrm{W\cdot m^{-1}\cdot K^{-1}}$, while the experimental value is close to 2000 $\mathrm{W\cdot m^{-1}\cdot K^{-1}}$. This is mainly because we did not account for the boundary scattering, while the Debye temperature of diamond is very high ($\sim2000\ \mathrm{K}$), meaning that 300 K can even be considered as a low temperature. Besides, the difference of isotope composition between experiment and calculation also accounts for such a difference of the thermal conductivity at low temperatures. At higher temperatures ($> 500\ \mathrm{K}$), the difference between our calculation results and the reference values is within $10\%$.

The ratios of the acoustic specific heat to total specific heat ($R_A$) and of the optical specific heat to total specific heat ($R_O$), which can reflect the relative contribution of the acoustic branches and optical branches, are shown in Fig. 2c. Within the temperature range of $\mathrm{200\ K - 1000\ K}$, $R_A$ is larger than $R_O$, indicating the domination of the acoustic branches. However, $R_A$ diminishes while $R_O$ grows as the temperature increases, making the contributions from the acoustic branches and optical branches almost equal above 1000 K.

Fig. 2d shows the phonon relaxation times in the resistive scattering processes, i.e. Umklapp scattering and isotope scattering, for each branch. From these data, we can compare the contribution of different scattering processes and understand the origin of thermal resistivity. For example, in the isotope scattering process, the relaxation times are 5.70E-11 $\mathrm{s}$, 4.64E-11 $\mathrm{s}$ and 9.47E-11 $\mathrm{s}$ for the $TA$, $TA’$ and $LA$ branches, respectively. In the Umklapp process, the relaxation times above 300 K are shorter than those in the isotope scattering process for all the acoustic branches, suggesting that the Umklapp scattering is stronger than isotope scattering. The relaxation times for optical branches are shorter than those for the acoustic branches by three orders of magnitude, suggesting a very strong scattering of phonons. 

\begin{figure}[htb]
\centering
\includegraphics[width=0.4\textwidth]{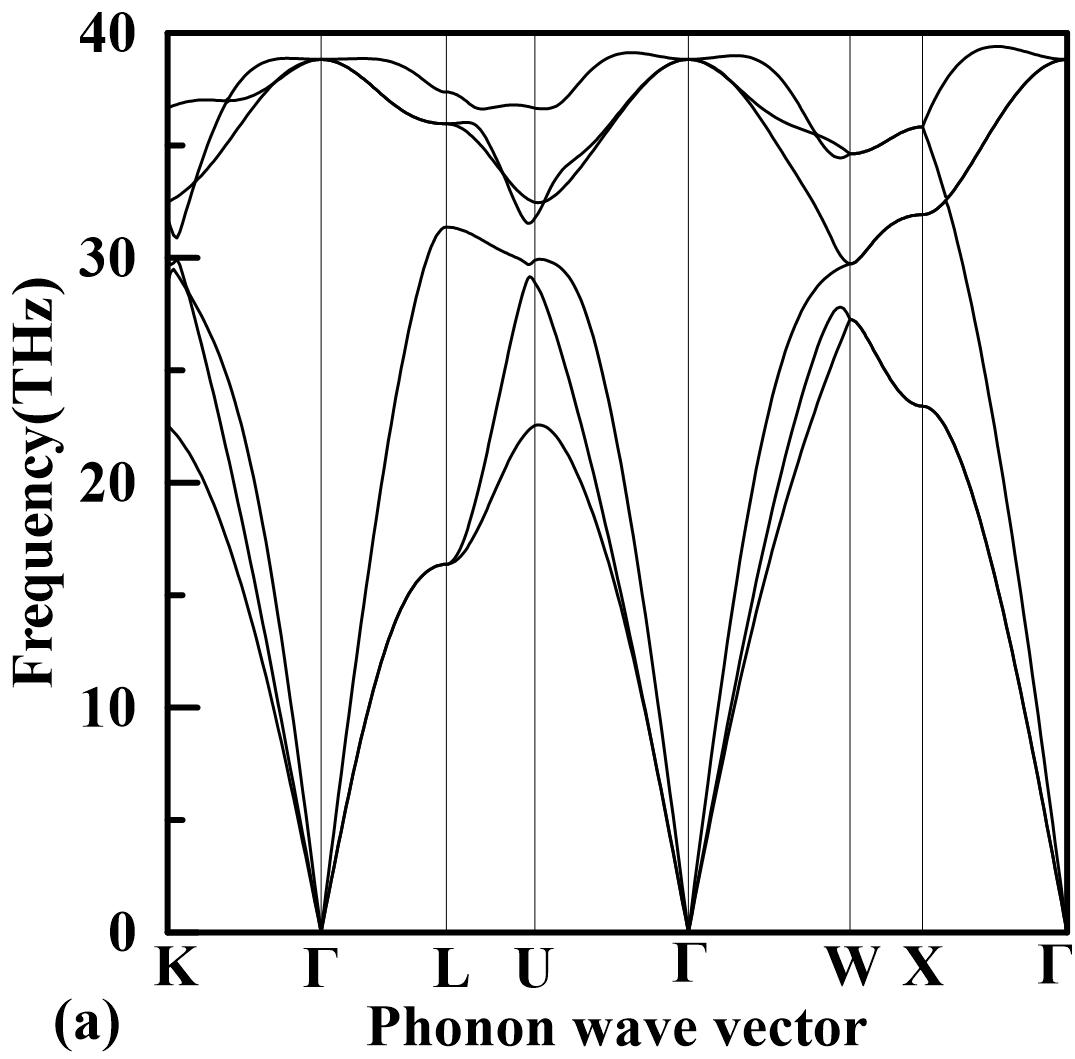} %
\includegraphics[width=0.425\textwidth]{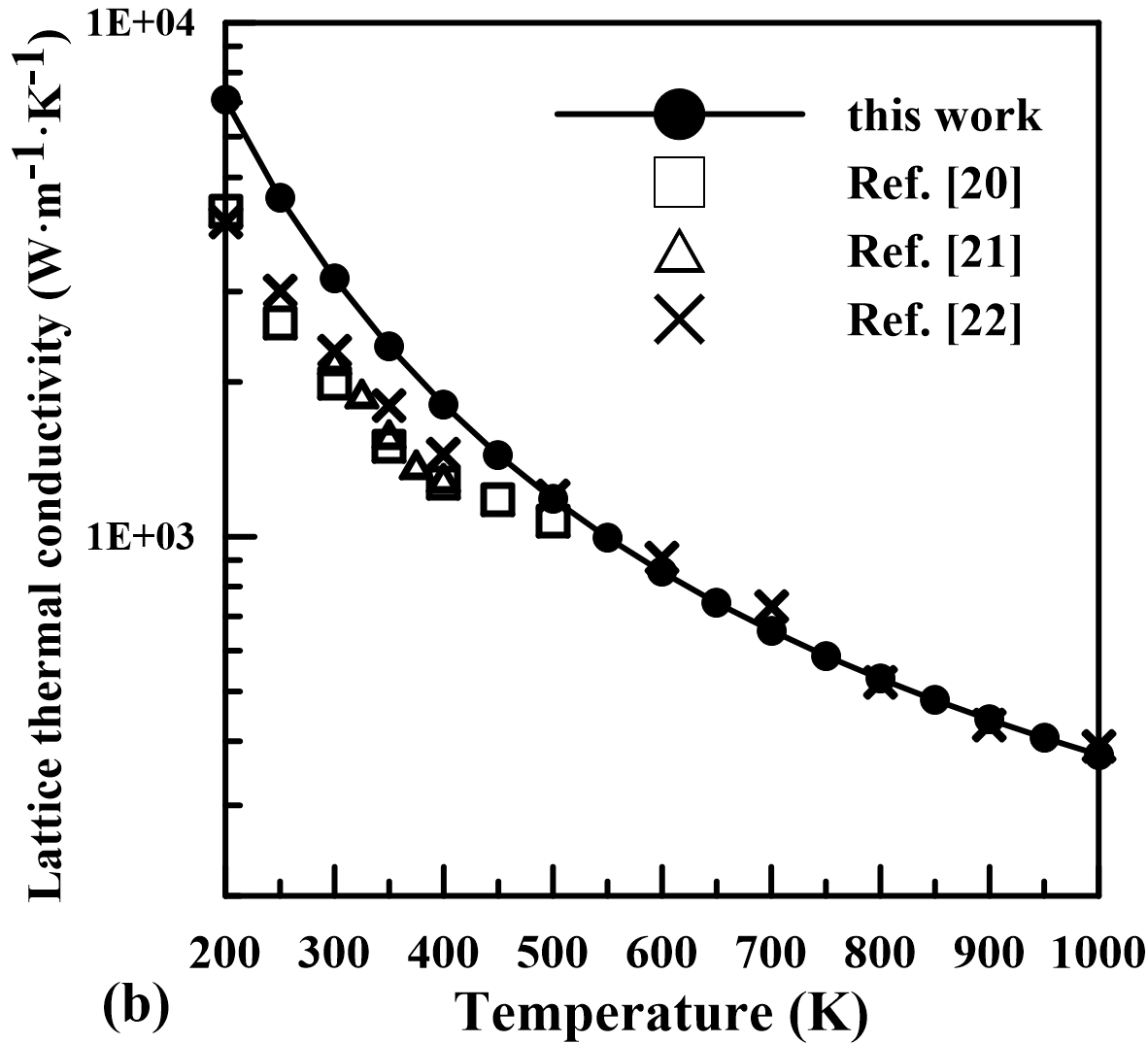} \\
\includegraphics[width=0.42\textwidth]{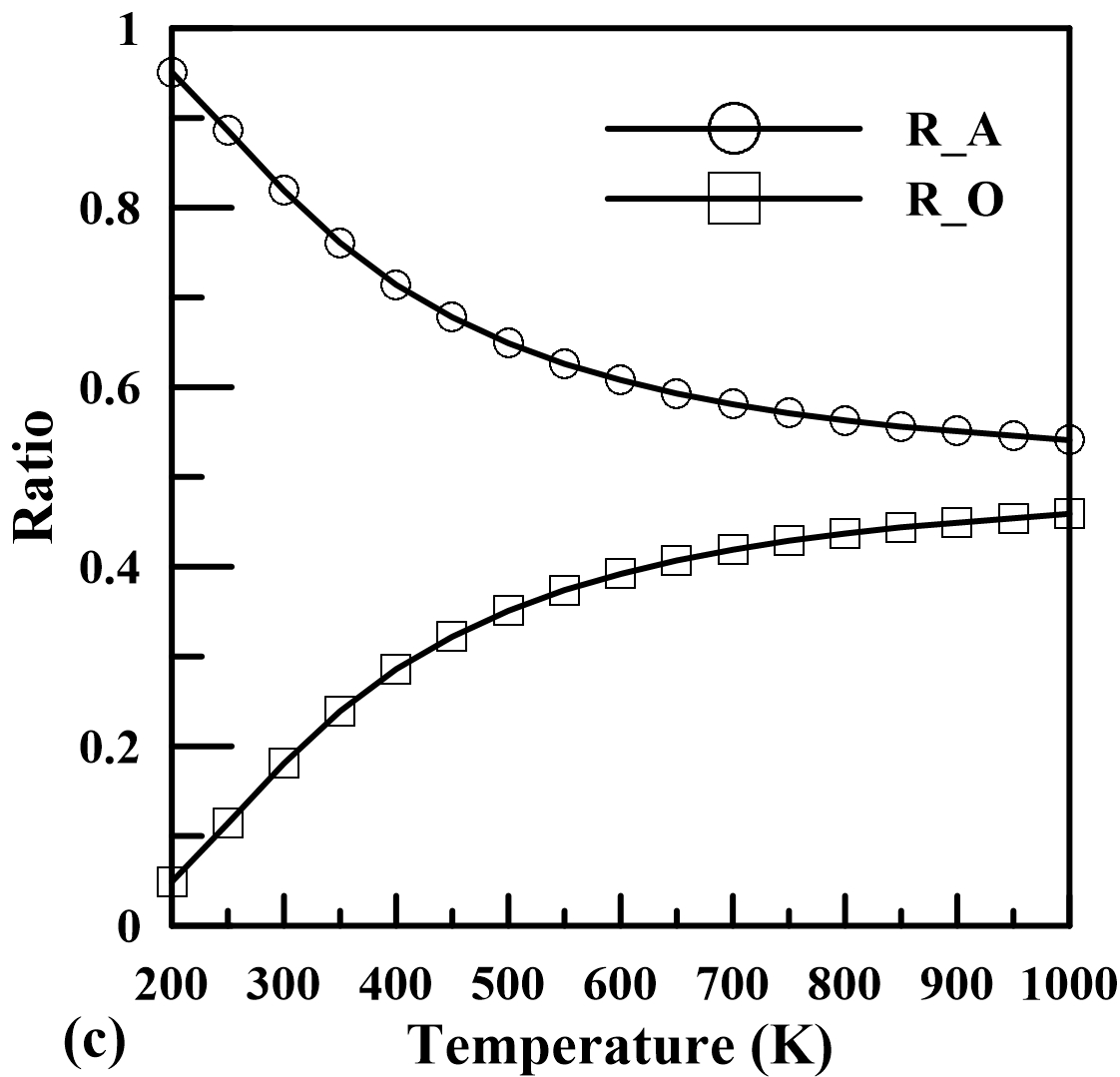}%
\includegraphics[width=0.435\textwidth]{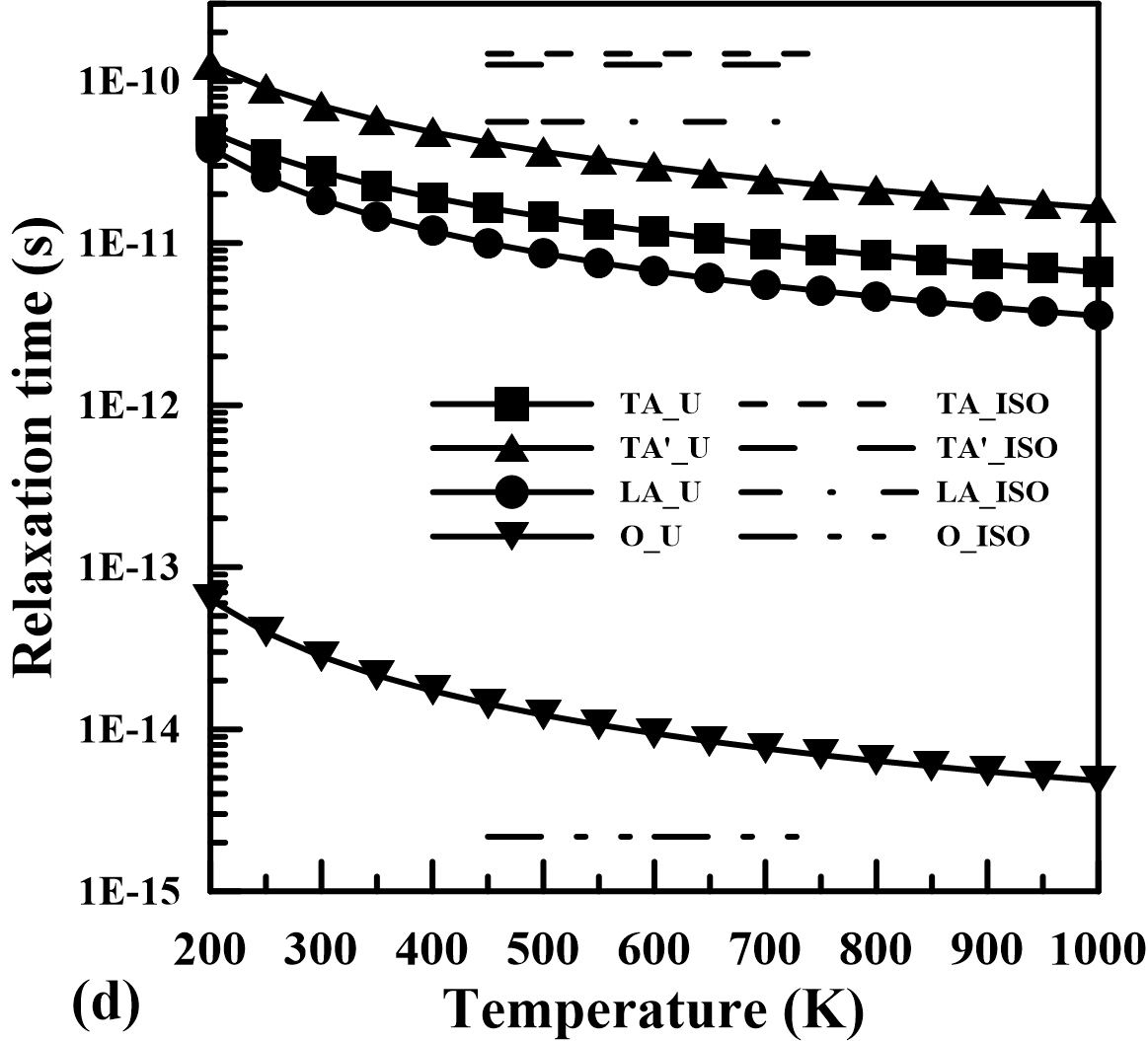}

\caption{(a) Phonon spectrum of diamond. (b) Lattice thermal conductivity of diamond calculated by our method compared with the experimental and full \textit{ab initio} results. (c) Percentage of specific heat from the acoustic and optical branches in total specific heat of diamond. (d) Phonon relaxation times for the resistive scattering processes.}\label{fig:2}
\end{figure}

\subsection{Silicon}

The calculated phonon spectrum and lattice thermal conductivity of Si are shown in Fig. 3a and 3b, respectively, the latter providing a comparison with the experimental results for the natural Si isotope\cite{kremer2004,touloukian1970}. The results are in close agreement with the experimentally measured values at temperature above 200 K. For example, at 300 K, the lattice thermal conductivity calculated using our method is 130 $\mathrm{W\cdot m^{-1}\cdot K^{-1}}$, while the experimental value is $140 \ \mathrm{W\cdot m^{-1}\cdot K^{-1}} - 143 \ \mathrm{W\cdot m^{-1}\cdot K^{-1}}$, which shows the difference of  $7\% - 9\%$. For temperatures below 200 K, the difference may be larger due to omission of the boundary scattering in the calculations and the difference of the isotope composition used.

The ratio of the acoustic specific heat to total specific heat, $R_A$, and that of the optical specific heat to total specific heat, $R_O$, are shown in Fig. 3c. The two ratios converge as the temperature increases. Similar to diamond, in the temperature range of $\mathrm{200\ K - 1000\ K}$, $R_A$ is larger than $R_O$, indicating the dominance of the acoustic branches. The phonon relaxation times in the resistive scattering processes are shown in Fig. 3d.

\begin{figure}[htb]
\centering
\includegraphics[width=0.4\textwidth]{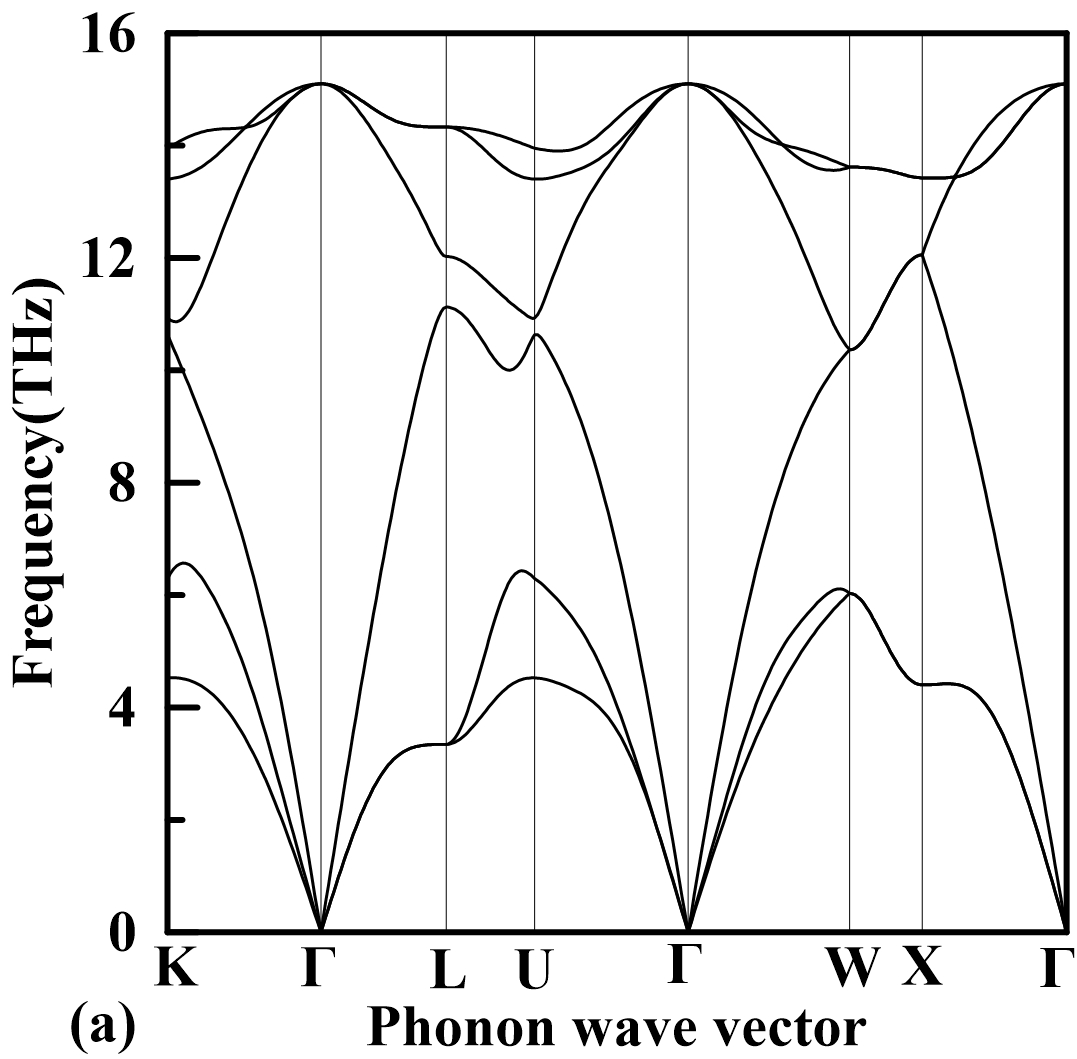} %
\includegraphics[width=0.41\textwidth]{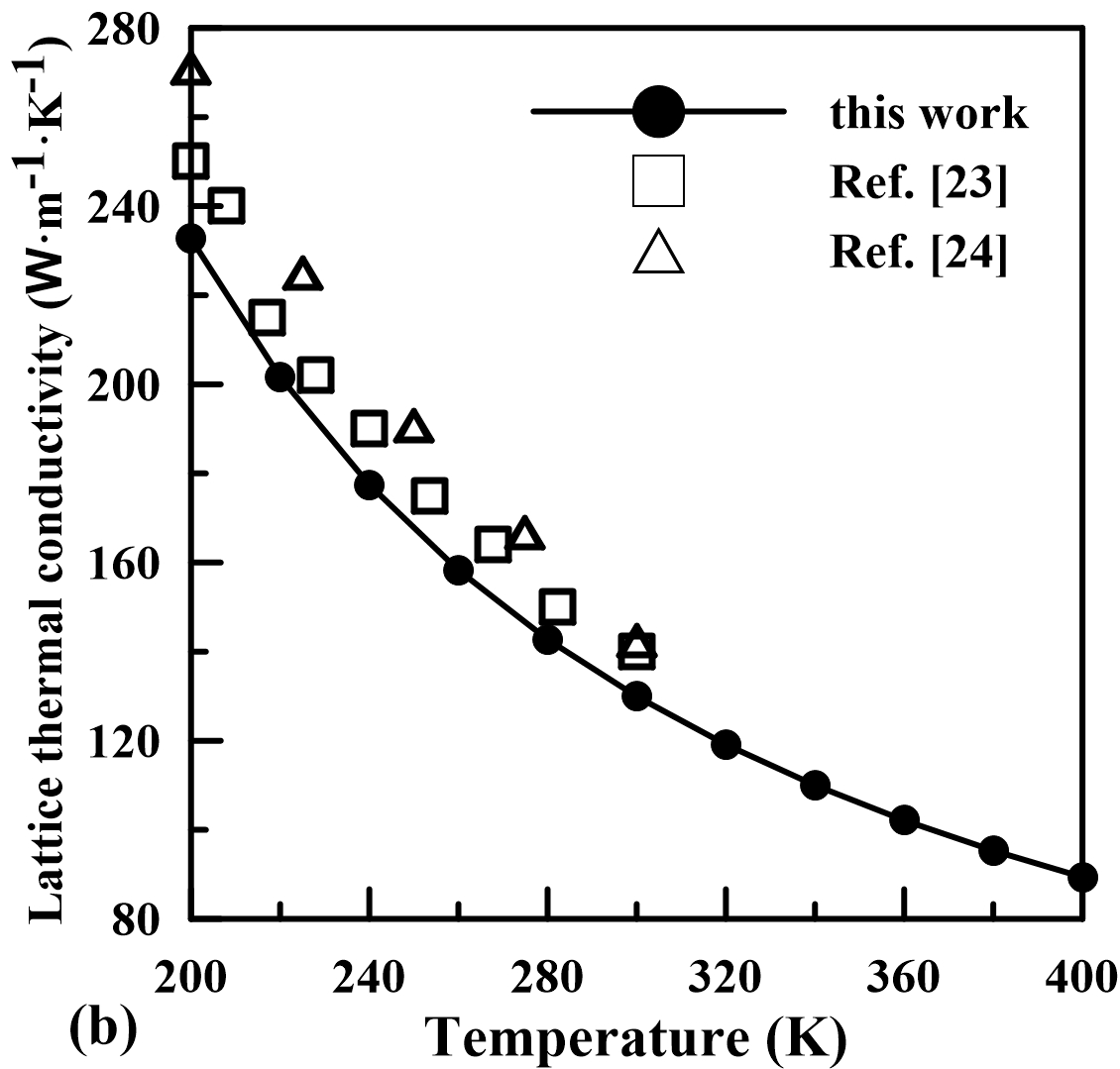} \\
\includegraphics[width=0.42\textwidth]{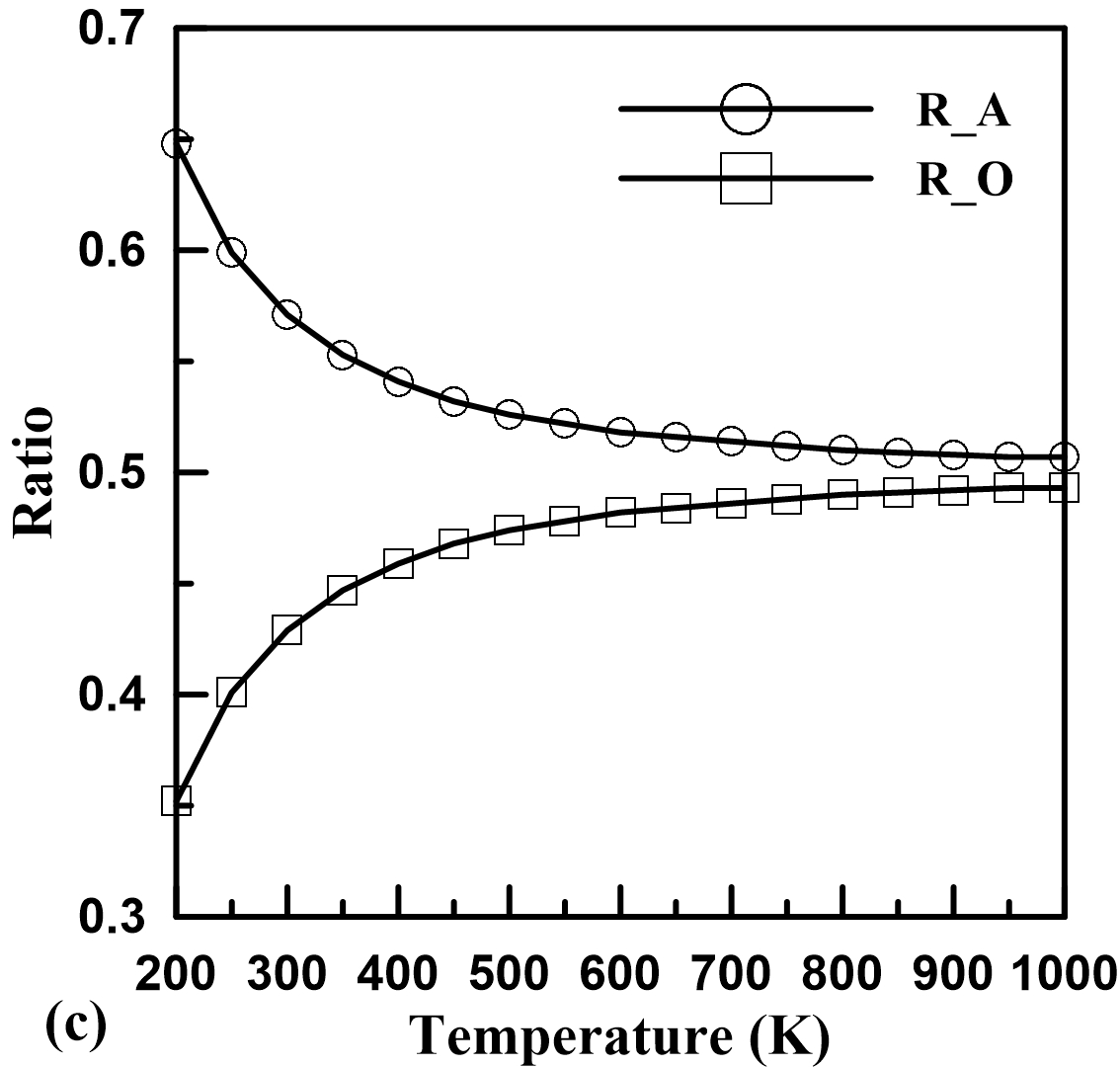}%
\includegraphics[width=0.435\textwidth]{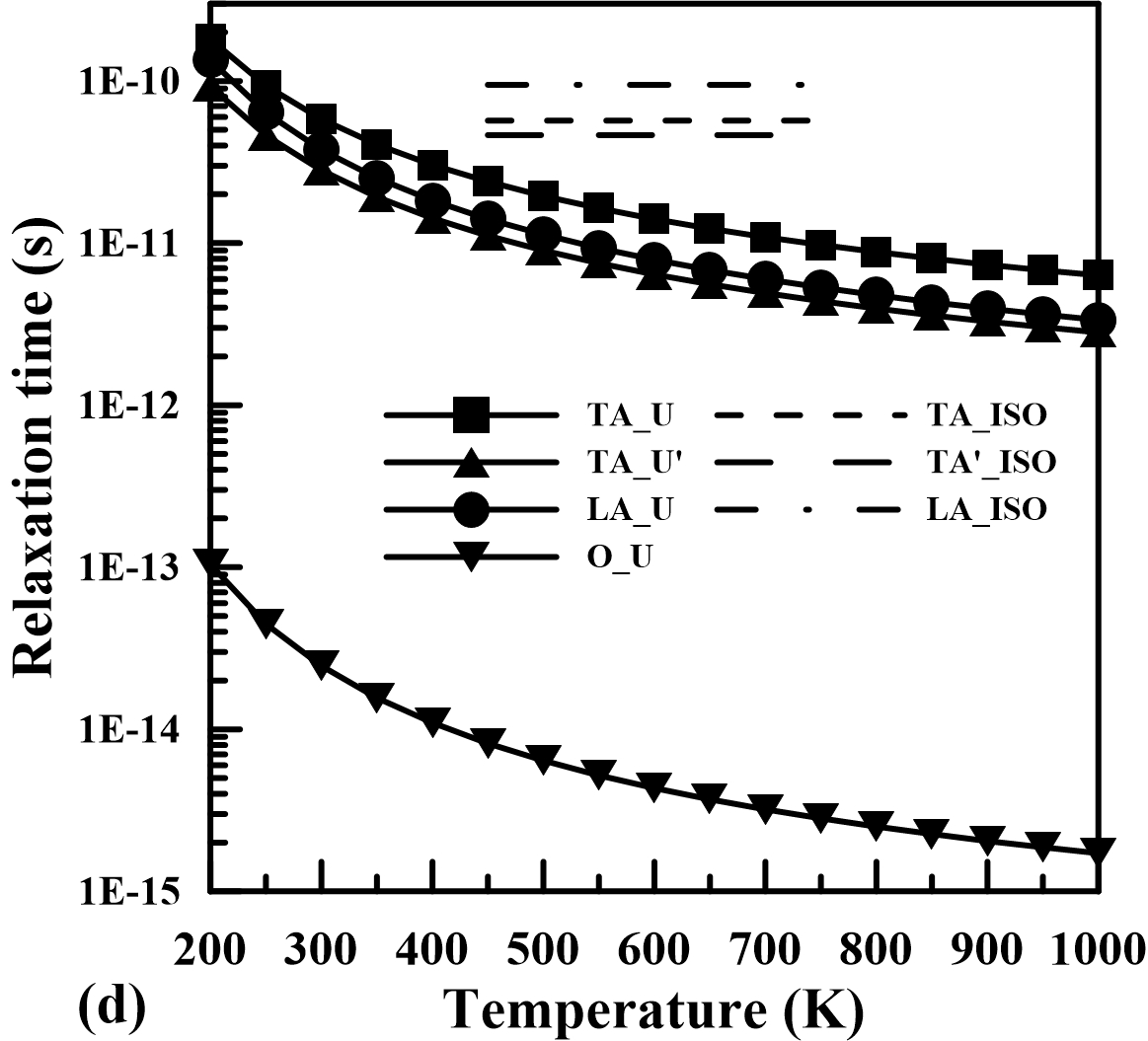}

\caption{(a) Phonon spectrum of Si. (b) Lattice thermal conductivity of Si calculated by our method, and compared with two experimental results. (c) Percentage of specific heat from the acoustic and optical branches in total specific heat of Si. (d) Phonon relaxation times for resistive scattering processes.}\label{fig:3}
\end{figure}

\subsection{Tin selenide (SnSe)}

SnSe is a newly discovered thermoelectric material with a very high figure of merit (2.6 at 923 K along $b$ axis)\cite{zhao2014}. Its crystal structure is highly anisotropic and its good thermoelectric performance is mainly due to its intrinsically very low lattice thermal conductivity along $b$ and $c$ axis. At about 750 K, SnSe goes through a phase transition from the low-temperature \textit{Pnma} structure to high-temperature \textit{Cmcm} structure. The high figure of merit values appear near and above the transition temperature. Here we calculated the lattice thermal conductivity of the low-temperature phase and compared the results with the experimental measurements.

The phonon spectrum and lattice thermal conductivity of SnSe are shown in Fig. 4a and 4b, respectively. The calculated $\kappa_L$ of SnSe obtained using our method is more than twice the experimentally measured values in the temperature range of 300 K to 1000 K. Nevertheless, our results still verify the fact that SnSe single crystal has a very low lattice thermal conductivity (around 1 $\mathrm{W\cdot m^{-1}\cdot K^{-1}}$), the reason being a low phonon velocity and high Grüneisen parameter, especially for the $TA$ branch (as shown in Table 1). In addition, an extremely low Debye temperature of the acoustic branches (Table 1) makes its specific heat converge quickly to the steady value, making the optical branch dominates the thermal conductivity at very low temperature (Fig. 4c), which is quite different from those of diamond and Si. This may suggest a possible way to find materials with low lattice thermal conductivity by looking for compounds whose primitive cell contains more than two atoms and whose Debye temperature from acoustic branches is very low. 

\begin{figure}[htb]
\centering
\includegraphics[width=0.4\textwidth]{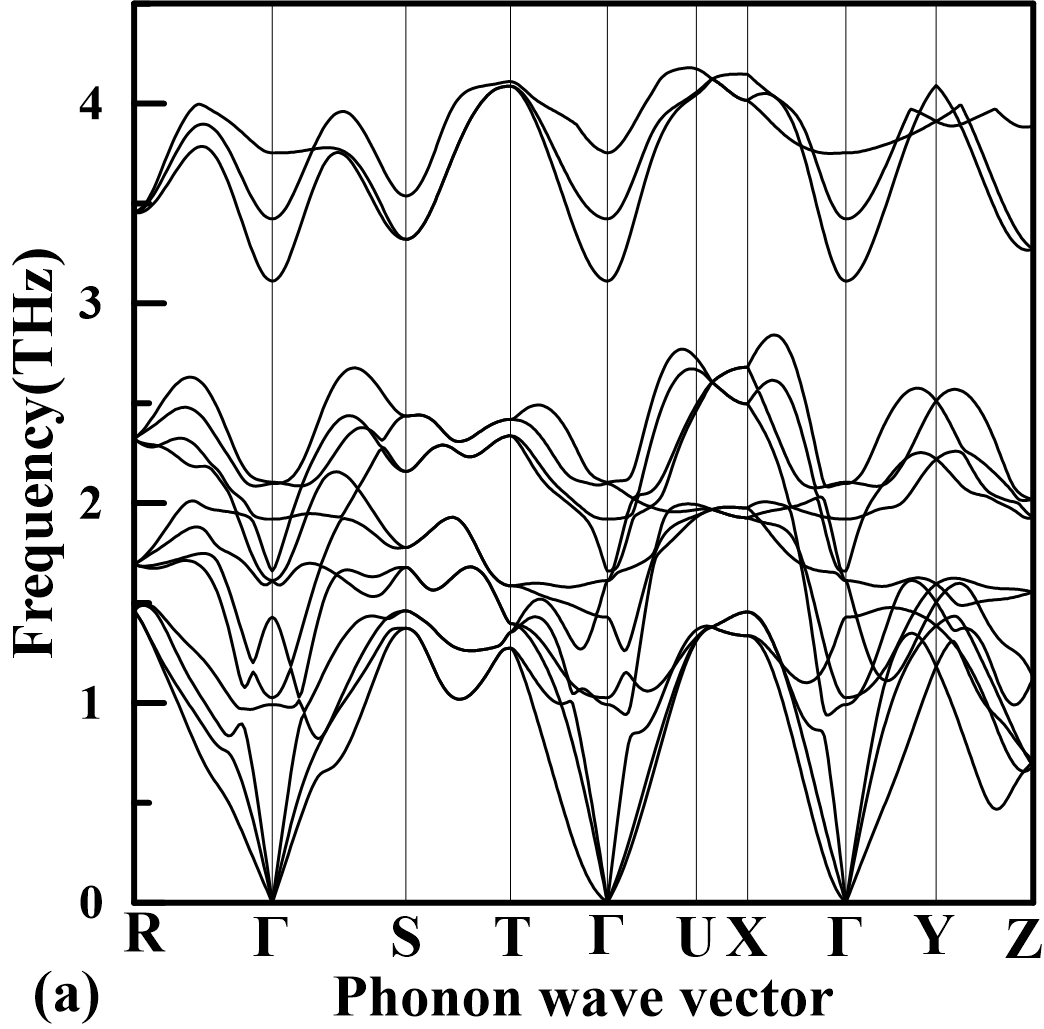} %
\includegraphics[width=0.42\textwidth]{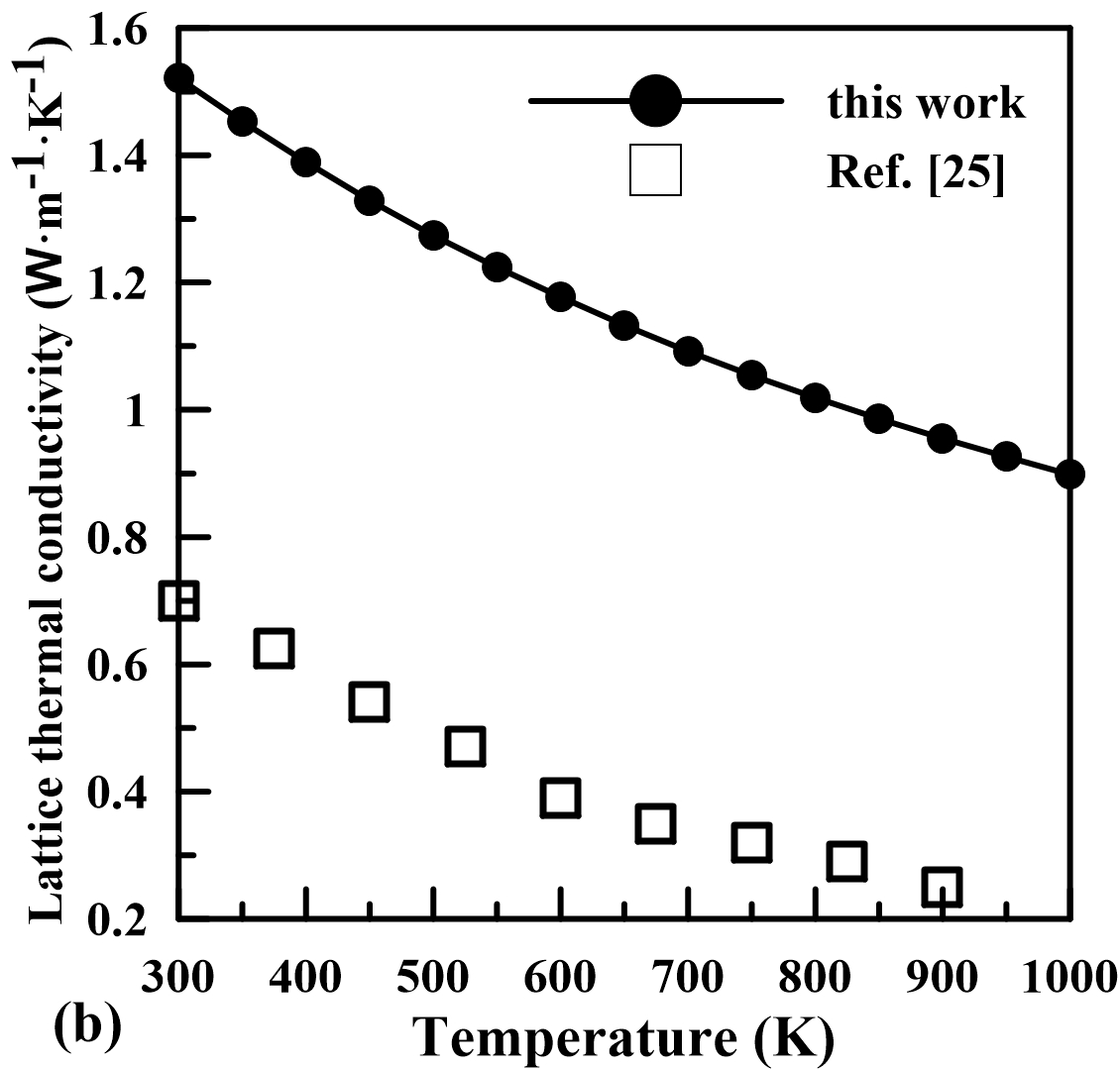} \\
\includegraphics[width=0.43\textwidth]{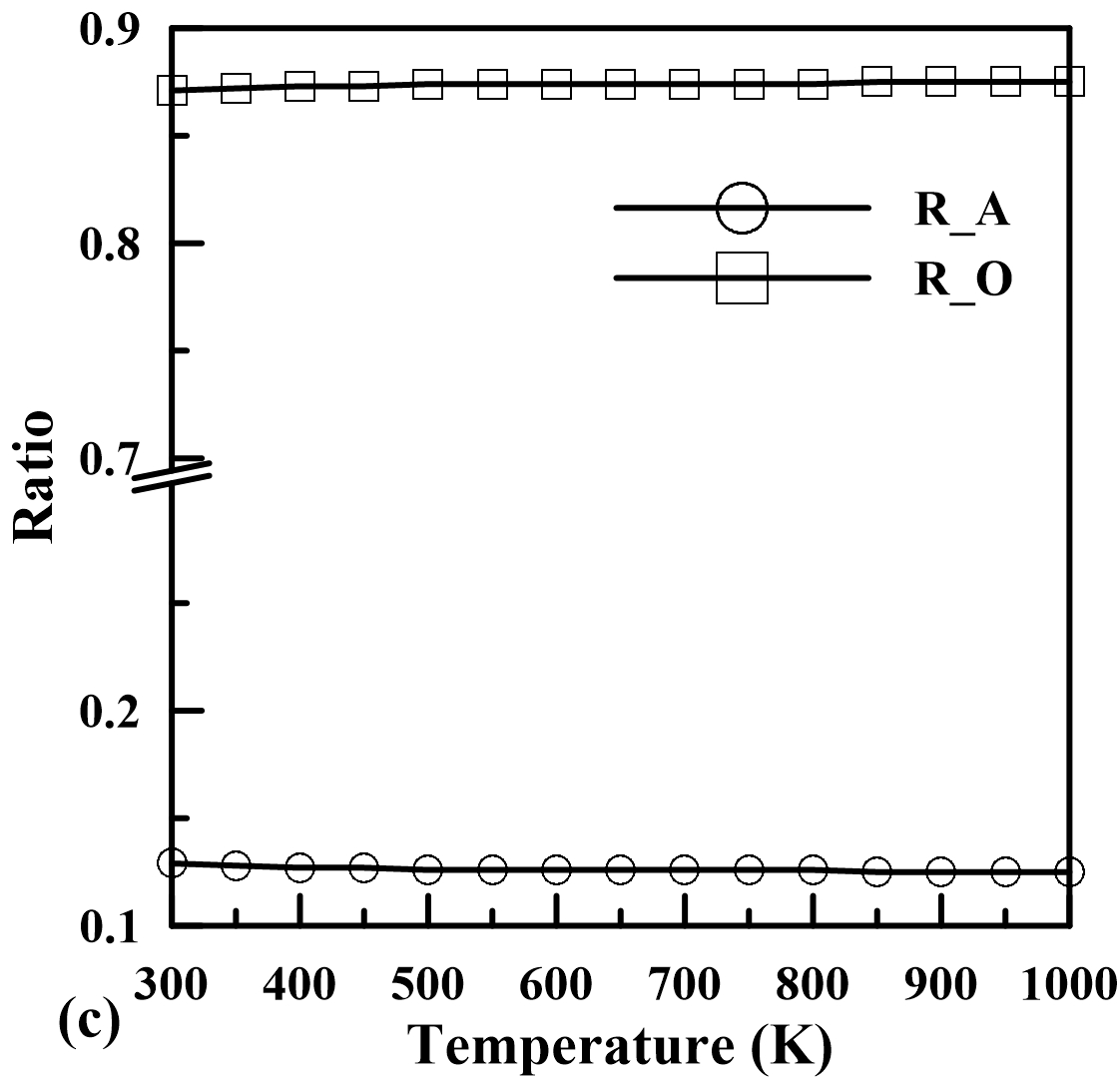}%

\caption{(a) Phonon spectrum of SnSe. (b) Lattice thermal conductivity of SnSe calculated by our method in comparison with the experimental results. (c)  Percentage of specific heat from the acoustic and optical branches in total specific heat of SnSe.}\label{fig:4}
\end{figure}

In general, the method proposed in this work can give reliable results for compounds with normal and high lattice thermal conductivity (dozens of $\mathrm{W\cdot m^{-1}\cdot K^{-1}}$ and above) in high temperature range ($> 200\ \mathrm{K}$). However, the values may be overestimated for compounds with very low lattice thermal conductivity (several $\mathrm{W\cdot m^{-1}\cdot K^{-1}}$), even at high temperatures. Even so, the results are meaningful, comparable, and relatively easy to obtain. 

\begin{table}[htbp]
\setlength{\tabcolsep}{0.1cm}
\caption{The average Debye temperatures ($\theta$), phonon velocities ($\nu$) and Grüneisen parameters ($\gamma$) of the transverse, longitudinal acoustic branches and optical branch in Si, diamond and SnSe.}\label{table:1}
\begin{tabular}{@{}*{13}{c}@{}}
\Xhline{2pt}
Material   &   $\thead{\theta_{TA}\\(K)}$   &   $\thead{\theta_{TA'}\\(K)}$   &   $\thead{\theta_{LA}\\(K)}$   &   $\thead{\theta_O\\(K)}$   &   $\thead{\nu_{TA}\\(m \cdot s^{-1})}$   &   %
$\thead{\nu_{TA'}\\(m \cdot s^{-1})}$   &   $\thead{\nu_{LA}\\(m \cdot s^{-1})}$    &   $\thead{\nu_{O}\\(m \cdot s^{-1})}$   &   $\gamma_{TA}$   &   $\gamma_{TA'}$   &   $\gamma_{LA}$   &   $\gamma_{O}$    \\
\Xhline{1pt}
Si   &   $289.4$   &   $315.3$   &   $576.9$   &   $723.2$   &   $4635.7$   &   $4931.3$   &   $8424.0$   &   $385.2$   &   $0.57$   &   $0.37$   &%
$1.04$   &   $1.18$                 \\
C   &   $1308.5$   &   $1415.5$   &   $1561.8$   &   $1872.7$   &   $11803.0$   &   $12236.6$   &   $17694.1$   &   $423.8$   &   $0.54$   &   $0.82$   &%
$1.05$   &   $1.07$                 \\
SnSe   &   $70.2$   &   $70.2$   &   $71.4$   &   $197.3$   &   $2212.5$   &   $2194.7$   &   $2948.9$   &   $303.7$   &   $16.71$   &   $3.32$   &%
$2.96$   &   $0.74$                 \\  
\Xhline{2pt}
\end{tabular}
\end{table}

\section{Conclusion}

We have presented a computer program mDCThermalC for calculating the lattice thermal conductivity using the modified Debye-Callaway model. Going beyond the original Callaway model, we take into consideration the optical branches, using the specific heat ratio as a weight to sum the contribution of the acoustic and optical branches. Our method take into accout the phonon-phonon normal scattering, Umklapp scattering, and isotope scattering processes, while all the necessary parameters can be calculated using first-principles methods. The capability of our program has been demonstrated on three examples of isotropic and anisotropic systems, which included compounds with very high and very low thermal conductivity. All the reviewed cases validate the robustness and accuracy of our method.

Compared to fully \textit{ab initio} program like ShengBTE, our program works much faster with the total calculation time approximately equal to that of a phonon spectrum calculation, since we do not need to calculate the anharmonical force constant. Our approach can be used for a wide range of materials and gives more accurate results compared with the original Callaway model or other semi-empirical methods. For different scattering processes, mDCThermalC directly calculates the estimated phonon relaxation times, providing additional information for studying the source of thermal resistivity.

One of the shortcomings of the presented approach is that the optical branch is treated as a longitudinal acoustic branch, which is clearly a rough approximation. In reality, the optical branch has its own dependence on the frequency and temperature which needs a further theoretical investigation to find the proper formalism. The other issue comes from the original Debye-Callaway model, where Callaway used the classical Boltzmann distribution to describe the phonon behavior. However, the Bose-Einstein distribution should be used for phonons, and the resulting equations may differ from the original ones. This has been addressed by Allen\cite{allen2013} and the modified formulas will be included in the future version of the presented method. We strongly encourage those who have similar research interests to join us and improve this program.

Overall, our program will be quite useful for a high-throughput screening of the lattice thermal conductivity of materials.

\section*{Acknowledgements}

The Author acknowledge the usage of the Skoltech CDISE HPC cluster PARDUS for obtaining the results presented in this paper. This work is funded by Russian Foundation for Basic Research (grant 18-29-12128/18).

\section*{Declarations of interest: none}





\bibliography{mDCThermalC}







\end{document}